# PROPOSED NEW DYNAMIC POWER INSERTION METHOD FOR STABILIZED POWER GENERATION BASED ON BATTERY ENERGY STORAGE SYSTEM

Amirhossein Khosravipour[1]

[1]Department of Electrical Engineering, Islamic Azad University Kermanshah Branch, Kermanshah, Iran
ah.khosravipour@iauksh.ac.ir

## ABSTRACT

*The solar energy is clean and future energy for electricity generation. Its energy has enormous potential but don't optimal to utilized caused by intermittent energy. The intermittent irradiance and ambient temperature influence the energy produced fluctuates and unstable. These power fluctuations affect system stability and frequency. To overcome this problem, several methods for PV power stabilization have been developed. One of them is the Stabilized Power Generation where PV output power is to a certain power value. The result of SPG method is reducing in PV power fluctuations but still unstable. For reaching the stable condition of PV Power output, the Dynamic Power Insertion method is proposed which is a modification of the SPG method with energy storage batteries. DPI improve the SPG method and make PV power stable with active power management with charge and discharge action. The battery is using as energy storage when PV power > Power limit. On the other hand battery as an energy source for active power insertion at PV power <Power limit. Thus the output power in each condition can be maintained as Power PV = P limit. For test this method, DPI modules are built on ARM lpc-1768 NXP and monitored with the Thingspeak webserver (simulated with Simulink/MatLab Software). The experimental results of DPI can stabilize PV power fluctuation at its setting power with an error of 5%.*

## 1. INTRODUCTION

Solar radiation in Iran is higher than the global average, about 1880 to 2250 kWh/m2 per year. Annually, more than 288 sunny days are recorded in more than 92% of Iran's lands, and this potentially produces a significant source of energy, due to global warming in recent years, this energy is increasing.

Iranians has the potential of solar energy for the technical generation of electrical energy which is equal to14.7 Twp but is only utilized at 0.021% or 309 Mwp(PV equivalent area 0.087% ).But behind that, some challenges must be overcome. This is due to the intermittent nature of Photovoltaic (PV), which is the volatility of PV output power which is fluctuatively influenced by the level of irradiation of solar radiation and environmental temperature. At one time the sun's energy can be maximal but can suddenly drop because of shading due to the movement of clouds [10].For energy intermittent problems and power instability, efforts have been made to mitigate PV power fluctuations using the method of hybrid PV and Diesel Generator[10,11].However, this method is less satisfying because in addition to being expensive and causing greenhouse gas emissions, also because of the ramp rate of the Genset it is not able to keep up with the momentary ramp down of PV power. Other methods include Power Curtailment, Energy Storage, Geographical Dispersion, Load Shaping and Power Smoothing [10, 11]. The Stabilized Power Generation method is a power curtailment method. Freede Blabjerg [5] provided a comprehensive explanation of the SPG method. The main purpose of the SPG method is to get a stable PV output with a certain power limit value. Illustration of SPG or power curtailment as shown below. Figure A is a portrait of PV power before SPG, while condition B is a portrait of PV power after SPG. With SPG the output power of PV is limited to its limit power and the PV output power in condition B is more stable than the output power of the PV condition A.

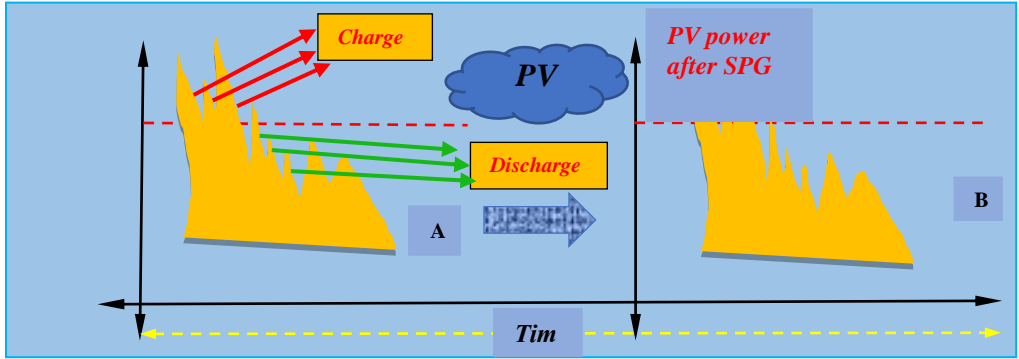

Fig. 1.(a) SPG concept (b). DPI concept.

To get a stable power output according to its power limit, F Blabjerg made modifications to the MPPT so that the PV worked in the MPPT operating mode and SPG mode. MPPT operations when PV power is less than a power limit while SPG operations when PV power is <limit power. Simple mathematically [5] can be expressed with:

$$P_{PV} = \begin{cases} P_{MPPT} \text{ when } P_{PV} \leq P_{LIMIT} \\ P_{LIMIT} \text{ when } P_{PV} > P_{LIMIT} \end{cases} \quad (1)$$

F Blabjerg [4], [5], [6] explained that there are three methods of SPG, namely P-SPG, I-SPG, and P & O SPG. PSPG is method with power-based regulation, I-SPG is method with current-based settings and P & O SPG is method by regulating PV voltage and MPPT voltage on the P & O algorithm.

The experiment conducted by Freede Blabjerg [4], [5], [6] is SPG without using energy storage. If SPG applied to PV output conditions in figure 1 condition B, the PV output power will be limited to its limit power. However, there is still a power valley in area B which fluctuates because the value is less than the limit power. To stabilize the PV output power at condition B it can be stable at its limit power, then the DPI method is proposed, namely the SPG method which is supported by energy storage batteries. Battery functions as energy storage when $P_{MPPT} > P_{LIMIT}$ and as an energy source for active power insertion at $P_{MPPT} < P_{LIMIT}$. Thus the output power in each condition $P_{PV} = P_{MPPT}$. DPI illustration as follows. Mathematically the DPI algorithm is written with the following equation [5]

$$P_{output} = \begin{cases} P_{PV} + P_{Battery} \text{ when } P_{PV} \leq P_{LIMIT} \\ P_{LIMIT} \rightarrow \text{when } P_{PV} > P_{LIMIT} \end{cases} \quad (2)$$

Where follows in eq (2).

$$P_{bat} = \begin{cases} P_{PV} - P_{LIMIT} \text{ when } P_{PV} \leq P_{LIMIT} \\ P_{LIMIT} - P_{PV} \text{ when } P_{PV} > P_{LIMIT} \end{cases} \quad (3)$$

To test this method a module is designed using ARM lpc-1768 accompanied by current, voltage, temperature sensors.

The results of DPI's performance in stabilizing PV power are monitored online at Thingspeak web server using the ESP 8266 device. The systematics of this paper consists of 4 parts. Part II describes the DPI methods and algorithms. Part III about designing modules and testing schemes and part IV analyzes the results of testing and conclusions. Based on the author's knowledge and observation, the design of the DPI module based on the SPG method using ARM lpc-1768 2560 is new. The results of DPI performance show that PV power fluctuations can be overcome and PV power output can be kept stable at its setting power.

*1.1 Structure and Operation DPI*

    A. DPI Structure

DPI is a prototype control device designed to mitigate intermittent effects and fluctuations in PV output power so that PV output power is stable according to its setting power. Figure 3 is a DPI configuration.

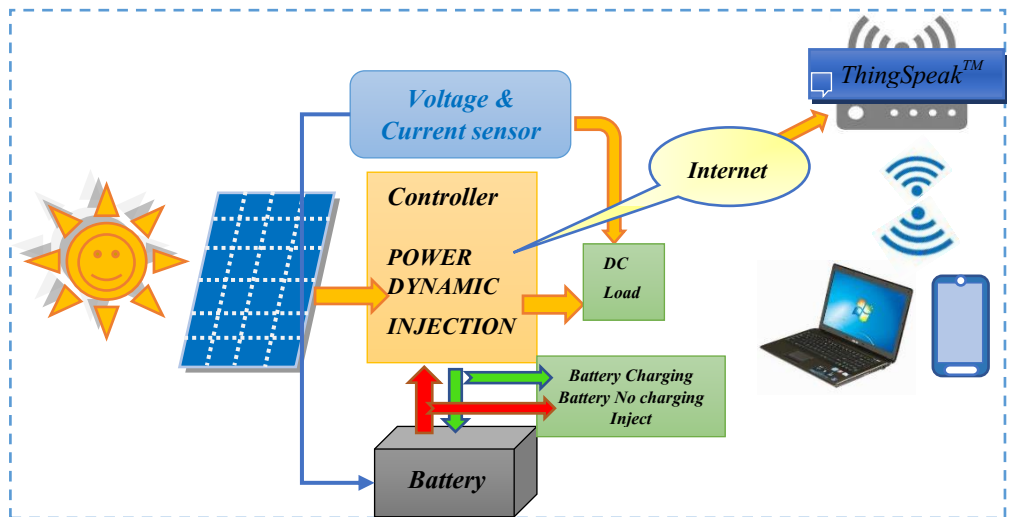

Fig. 3. Configures of DPI

By the fig.3, the DPI component consists of a controller module, current, voltage, temperature, power MOSFET, and battery sensors. DPI component equipment specifications are explained in section III. The description of the DPI controller referring to the diagram above can be described as follows:

i. Current and voltage sensors installed in the PV panel output will read the PV panel output voltage and current values. From reading current and voltage PV output power will be obtained.

ii. The control algorithm will measure the output power of PV and compare it with the setting power for PV output power settings. If the PV power exceeds the setting power, then the excess power will be saved to the battery. But if the PV output power is less than the setting power, then the active power stored in the battery will be inserted into the load.

iii. With dynamic insertion power control, PV power output stabilizes and intermittent problems can be resolved.

iv. DPI working parameter data and PV output power will be sent wirelessly using the Wi-Fi sensor module and done by cloud computing using the Internet of Things (IoT) application so that it can be monitored in real time on desktop PCs and Android phones.

B. DPI Algorithm

Based on the DPI concept in figure 3, a DPI algorithm is created by combining the SPG and battery methods with the following flow chart figure 4.

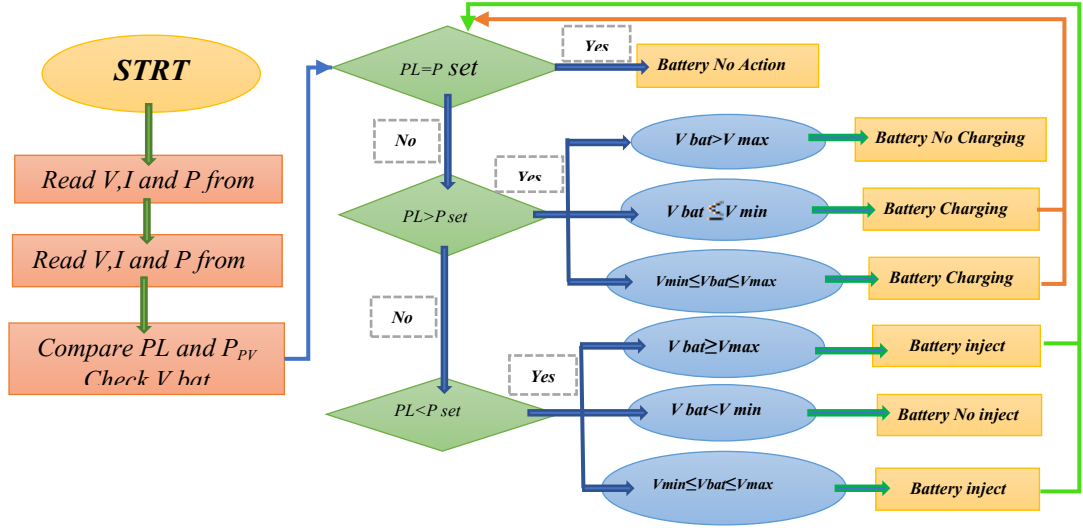

Fig. 4. Configures of DPI

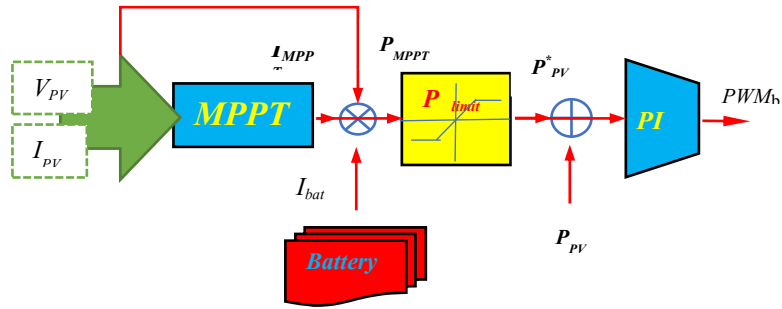

Fig. 5. DPI Control

The DPI method based on the P-SPG concept uses PI control as shown fig.5. Duty Cycle for PWM control is defined by a mathematic expression:

$$Duty = \frac{P_{error}}{P_{set}} \times 300 \text{ or } Duty = \frac{(P_{set} - P_{load})}{P_{set}} \times 300 \quad (4)$$

## 2. DPI DESIGN

The DPI module is built based on the ARM lpc-1768 Mega 2560 microcontroller which is then embedded in this controller programming algorithm to stabilize PV output power. The DPI component is detailed in table 1.

Table 1. DPI module hardware specifications.

| No | Tools | Specification |
|---|---|---|
| 1 | Keypad | Keypad 3 * 4 |
| 2 | ARM lpc-1768 Mega/NXP | Mega 2560, 16 MHz, 54 pins I/O, 5 Volt, 256 KB |
| 3 | Temperature Sensor | DHT 11, 50 C+ 2 |
| 4 | Voltage Sensor | 25 Volt DC |
| 5 | Current Sensor | WCS 1800, 35 Amp, sensitivity 60 mA/V |
| 6 | Power MOSFET | IRF5305S, 22Amp, 5 -12 V DC, opto coupler |
| 7 | LCD | LCD crystal 4x 20, baud rate 9600 |
| 8 | ESP 8266 v.1 | IEEE 802.11 b/g/n, BSS Station, 64 KB instruction RAM, 96 KB data RAM, 64 KB ROM and 1 MB Flash, TCP IP Stack, baud rate 115200 |

To conduct DPI performance testing additional tools are shown in table 2 below.

Table 2. Additional tools for DPI testing

| NO | Tools | Specification |
|---|---|---|
| 1 | Modul PV | single Crystal 100 Wp |
| 2 | SCC | SCC PWM 10 Amp |
| 3 | Battery VLRA | Battery SABA 42 Ah/12 Volt |
| 4 | DC Load | DC Lamp 20 Watt |
| 5 | Measuring Equipment | Hioki Clamp Amp 40 A |

In this proposed scheme, DPI using 3 power MOSFET that connected to PV, battery, and load. Power mosfet~1 to charge action, power mosfet~3 as insertion action and power mosfet~2 as bypass control.

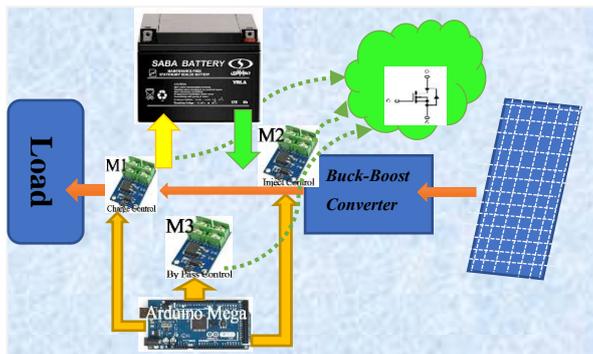

Fig. 6. DPI charge and discharge scheme.

## 2.1. EXPERIMENT RESULT AND ANALYSIS

### a. PV Monitoring Stabilization on Thingspeak

To find out the performance of the DPI module to stabilize PV power and its monitoring in real-time based on the Thing speak Webserver, the DPI was tested with P set = 13 Watt test parameters. In this test, two shading simulations were carried out by covering a portion of the PV panel. As a result of the shading simulation, the PV output power will decrease. DPI will work to stabilize PV power by inserting active power until setting power is reached. By using the WIFI ESP 8266 module and the *Thingspeak* web server application, it can monitor the parameters of PV power fluctuations, DPI performance, and battery status can be done wirelessly via PC and Android. The performance parameters monitored through the Thingspeak web server include Temperature, Setting Power, PV Power, Load Power, and Battery Power, as follows.

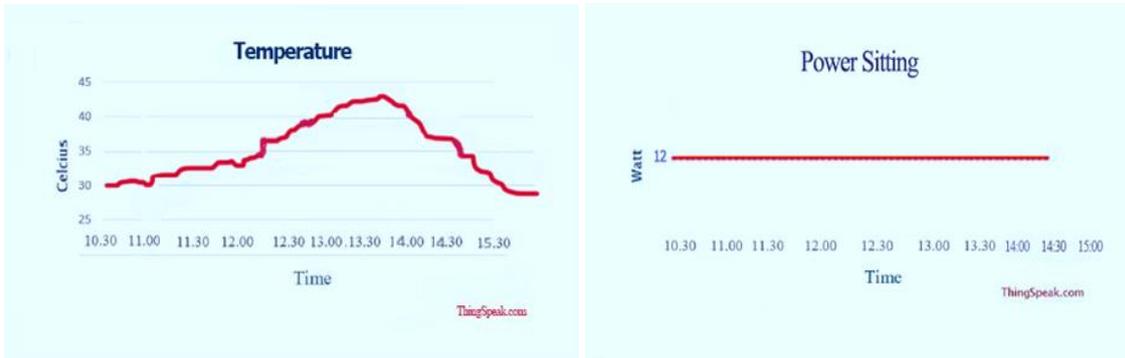

a. Temperature                                             b. Sitting power

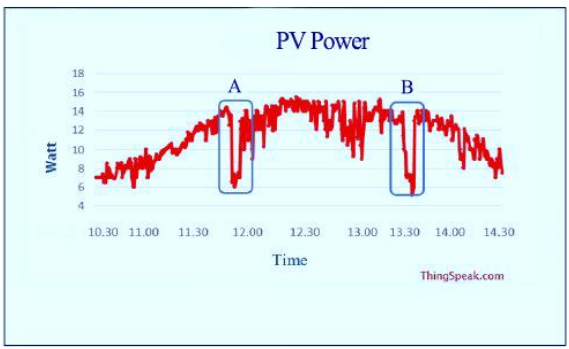                     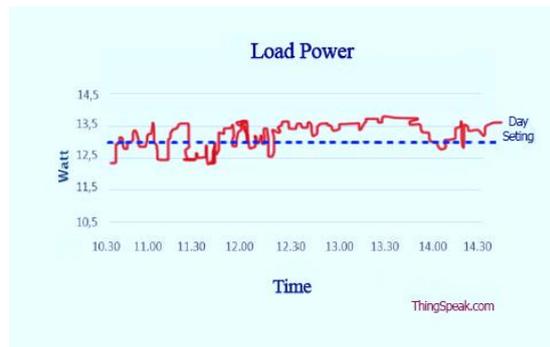

c. Monitoring PV Power                              d. Monitoring Load Power

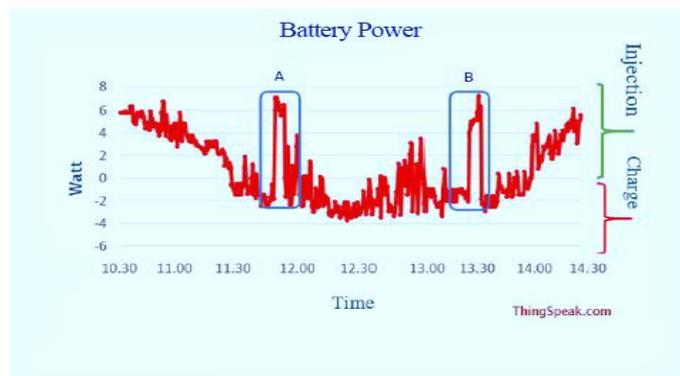

e. Monitoring Battery power

Fig. 7. (a) Temperature ; (b) Setting Power; (c) Monitoring PV Power ; (d) Monitoring Load Power ;(e) Monitoring Battery Power ;
(a ~ e) Monitoring by using *Thingspeak.*

   b.  *Power Stabilization Analysis with DPI*

To determine the performance of DPI in stabilizing PV power, it is necessary to compare several graphs of monitoring Results including PV power charts, battery power graphs, and load graphs. Merging these 3 graphs as below.

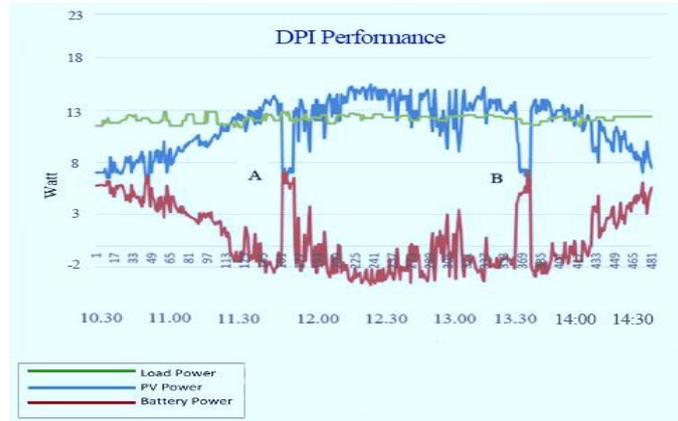

Fig. 7. DPI Performance.

Based on the graph above, it can be seen that the load power is in the range of 12 Watts with a deviation of + 0.5 Watt. Thus the DPI error rate is as follows:

$$Error = \frac{\sum (P_{cde} - P_{set})}{N} = 5,524\% \quad (5)$$

The fluctuations in the decline in PV power especially at points A and B are compensated by the DPI with the action of Insertion of battery power. Conversely, when there is a fluctuation in the increase in PV power, it is compensated by the DPI with absorbing action. Thus there is a balance between PV power and battery power so that a stable load power is Obtained according to the setting power. With the action of absorbing and insertion of the battery above, the rate is slim. One important DPI performance parameter is the slender response rate. The ramp rate is the response speed of adaptive Devices to mitigate the effects of disturbances to achieve setting values as soon as possible.

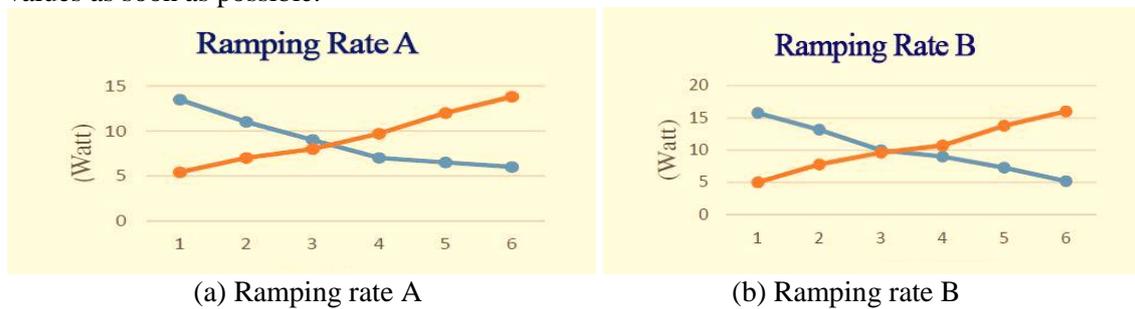

(a) Ramping rate A          (b) Ramping rate B

Fig. 7. Ramping Rate PV and Batteries.

From the graph above, the ramp rate of battery and PV are shown in table 3.

Table 3. Ramping rate Batteries and PV.

| Ramp Rate | PV | Batteries |
|---|---|---|
| A | 1,25 Watt/s | 1,4 Watt/s |
| B | 1,767 Watt/s | 1,83 Watt/s |

Based on Figure 5.c. and 5.d. then analysis and calculation using MatLab on energy generated by PV directly with the energy resulting from stabilization with DPI.

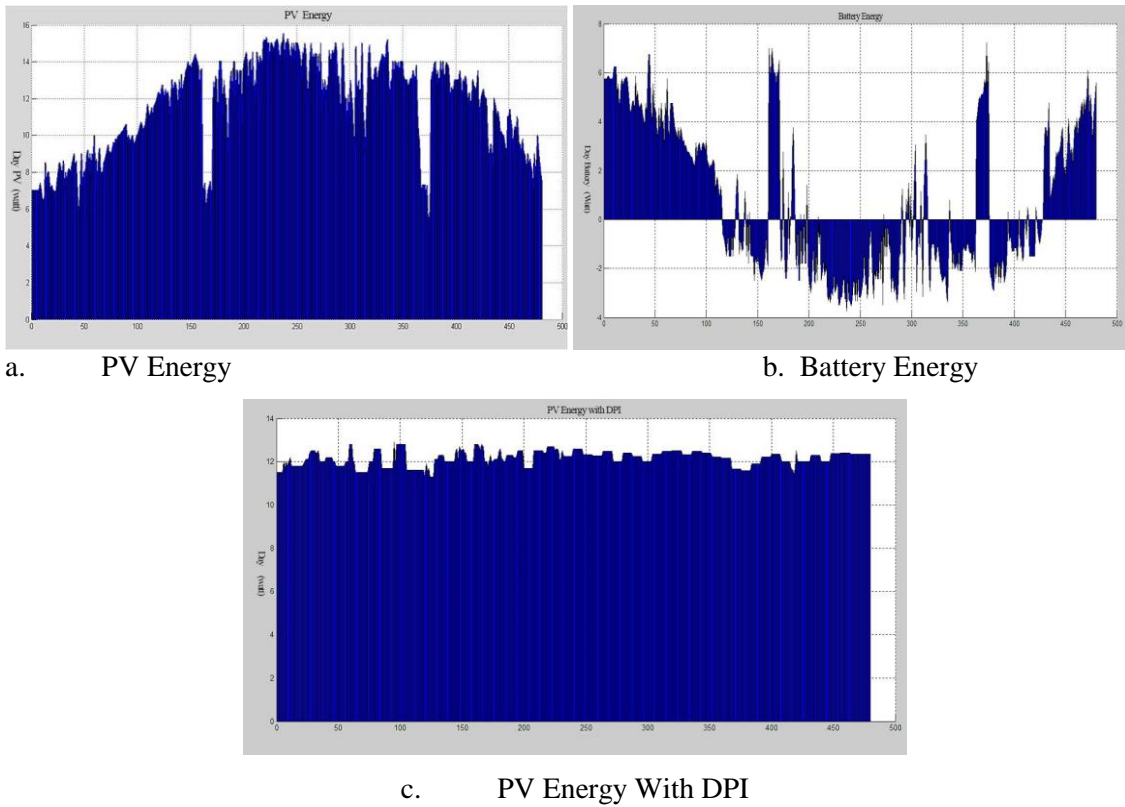

a.      PV Energy        b.  Battery Energy

c.      PV Energy With DPI

Fig. 8. Behavior energy diagram in MatLab.

From figure 8 (a) and (c) can be obtained that PV energy is 45,842 Wh. while the stabilization energy with DPI is 44,467 Wh. Thus there is a difference of 1,375 Wh. The difference is due to an error and a loss factor. But in general, the performance of DPI in regulating power stabilization due to PV fluctuations is excellent performance. Because the energy from the DPI arrangement is almost the same as the energy generated by PV.

## 3. CONCLUSIONS

DPI is a prototype module to stabilize PV power according to the setting power. The DPI method is a development of the SPG method combined with energy storage batteries. The DPI takes charge when PV Power bigger than P limit and discharging when PV Power smaller than P limit. Thus the PV output power can be kept stable at the setting value. The results of the DPI module with ARM lpc-1768 design drawings show the following results.

a) The DPI module error rate for stabilizing PV power is around 5%. For the 12 Watt setting power, the DPI output still fluctuates between 11.5 Watts and 12.5 Watts.

b) The DPI module works well with a ramp up battery that is greater than the ramp down PV.

c) Readings between serial monitors on the DPI with Thingspeak Webserver have a delay of 15 seconds.

**Authors**

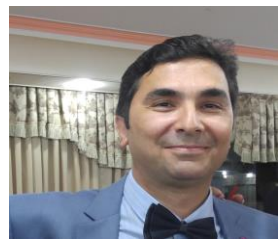

I was born in 1977 in a cultural family in Ilam, Iran. In 1995, I received a diploma in mathematics and physics from the state model high school and a bachelor's degree in applied physics from Razi State University in Kermanshah in 2000. I started working in 2001, I worked in Hyundai Design and Construction Company, Development of Phases 4,5 of South Pars Assaluyeh for 2 years. Expert of the engineering unit of the Ministry of Roads and Urban Development from 2005 to 2009. Master of Electrical Engineering Repairs in SRU of Ilam Petrochemical Company. Teaching different electrical engineering courses in universities. I got a master's degree in electrical engineering from Tehran University of Science and Research, Iran. Holds a doctorate in electrical engineering from Kermanshah University of Science and Research. Supervisor and secretary of Suggestion system of Ilam Petrochemical Complex Company since 1399 until now. Researcher. Publication of the book "Increasing the efficiency of brushless motors and examining optimization methods" in Persian (ISN: 987-622-7485-03-5) , with several articles in Persian and English presented in national and international conferences . Membership as scientific secretary, scientific committee and refereeing of several international conferences CIEE.IR, Tech.Sdcongress.ir, Choe.ir, IKT2019.IR, aeroconf.ir, eesconf.ir, icconf.ir emtc.ir, pedstc2020,14Sastech.khi.ac.ir,mhconf.ir , dmeconf.ir , dsconf.ir , 3icsce.ir , confitc.ir , ieedt.bcnf.ir , ren-c.ir , awpc2019.ir ,kbei.iust.ac.ir ,cnf.vru.ac.ir ,icietconf.com ,tetsconf.com,psc-ir.com,CEEG.org, ECEE.org, etc.